\documentstyle[preprint,12pt,prb,aps]{revtex}

\topmargin = - 1 cm
\begin{document}

\title{Phase transitions of electron-hole and unbalanced electron
systems in coupled quantum wells
 in high magnetic field}

\author{Yu.E.Lozovik\cite{email},
 O.L.Berman, V.G.Tsvetus \\
Institute of Spectroscopy, 142092 Troitsk,   \\
Moscow region, Russia }

%\date{}

\maketitle

\begin{abstract}
Superfluidity of spatially separated electrons and
holes and unbalanced two-layer electron system
in high magnetic field is considered.
 The temperature $T_{c}$ of the Kosterlitz-Thouless transition
 to a superfluid state    is obtained as a function of magnetic field
$H$ and interlayer separation $D$.
The equation of state for magnetoexciton
system in quasi-classical regime is analyzed.  The transition from
excitonic phase to electron-hole phase is considered.
 Possible experimental manifestations of
the predicted effects are briefly discussed.
\end{abstract}



\section{Introduction}

%\begin{large}

Systems of excitons with spatially separated electrons ($e$) and holes
 ($h$) (indirect excitons) in coupled  quantum wells (CQW) in
magnetic fields ($H$) are now the subject of
 intensive experimental investigations.
\cite{Zrenner}$^-$\cite{Timofeev} They are of  interest, in particular,
in connection with the possibility of superfluidity of indirect excitons
or $e-h$ pairs, which would manifest itself in the CQWs as {\it
persistent} electrical currents in each well  \cite{Ydson} (see also
recent articles \cite{Berman}$^,$\cite{Rice}$^,$\cite{Conti})
and also in connection with
quasi-Josephson phenomena in the system (see Ref.[\onlinecite{Klyuch}]
and references therein).  In high magnetic fields two-dimensional (2D)
 excitons survive in a substantially wider temperature region, as the
exciton binding energies increase with magnetic field.
\cite{Ler}$^-$\cite{Kallin} In addition, 2D $e-h$ system in high
 fields $H$ is interesting due to the existence, under some
conditions, supersymmetry in the system (for the single quantum well)
 leading to
 unique exact solutions of the many-body problem (the last corresponding
to ideal Bose-condensation of magnetoexciton at {\it any} density)
 \cite{Lerner}.

The superfluid state appears in the system under consideration below the
temperature of Kosterlitz-Thouless transition. \cite{Kosterlitz}  The
latter was studied recently for systems with spatially separated electrons
 ($e$) and holes ($h$) in the absence of magnetic field. \cite{Berman}

Attempts of experimental investigation of magnetoexciton superfluidity in
coupled quantum wells \cite{Zrenner} make it essential to study  the
 magnetic field dependence of the temperature of phase transition to the
superfluid state in systems of indirect magnetoexcitons   and to analyze
the density of the superfluid component. This is the subject of this
paper. It will be shown below, that increasing of magnetic field at a
fixed magnetoexciton density leads to  a lowering the
 Kosterlitz-Thouless transition temperature $T_{c}$ on account of the
 increase of the exciton  magnetic mass as a function of $H$.  But
it turns out that the  highest possible Kosterlitz-Thouless
transition temperature increases with $H$ (at small $D$) due
 to the rise in the maximum density of magnetoexcitons {\it vs.}  $H$.
Quantum phase transition of magnetoexciton system to incompressible liquid
 states is  briefly
 discussed in connection with the problem of maximum density of
stable magnetoexciton system.

We show also that contributions to the thermodynamic potential
 and the state equation connected with interexciton interactions for rare
magnetoexciton system at fixed density and temperature
vanishes with  $H$ increasing.

In magnetoexciton
system in coupled  quantum wells  at essentially higher
temperatures than the transition to the superfluid state the exciton
thermal ionization takes place.  The
 dependence of maximal $T_{i}(H,n)$
 is defined by the magnetoexciton binding energy
$E_{0}(H,D)$ and rises with magnetic field and decreases with
 interlayer separation $D$ ($n$ is the surface density of magnetoexcitons).
%The crystallization of indirect excitons was also studied \cite{Quinn},
%\cite{cr}.

The paper is organized in the following way.  In Sec.II (which has
auxiliary character) we discuss the
relation between electric current and magnetic momentum of noninteracting
isolated magnetoexcitons which we shall use in calculation of the density
of normal component.  In Sec.III we consider the spectrum of
collective excitations for the system of rare indirect excitons in high
magnetic field in the ladder approximation. In Sec.IV we analyze the
dependence of the density of the superfluid component   on magnetic field
and interlayer distance. In Sec.V we calculate the dependence of the
temperature of the Kosterlitz-Thouless
transition to the superfluid phase   on magnetic field and interlayer
distance.
In Sec.VI we
consider  thermodynamics and equation of the state of rare magnetoexciton
 system at high temperatures and discuss the resemblance of the system to
the ideal gas in the limiting case of high magnetic field.
 In Sec.VII we estimate the
magnetoexcitons existence line $T_{i}(H,n)$  connected
 with  ionization transition (more strictly, this is
 crossover region). We use
 the ionization equilibrium condition
analogous to Saha relation in the quasiclassical region  and
also  discuss briefly the quantum region of magnetoexciton system stability
in relation with the quantum transition of superfluid magnetoexciton system
to two-layer
Laughlin liquids of electrons
 and holes.  In Sec.VIII we discuss the
phase transitions in the "dense" system.
 In Sec.IX we consider properties of
indirect magnetoexcitons in unbalanced two-layer electron system. In
Sec.X we discuss possible experimental manifestations of superfluidity of
magnetoexcitons in CQW. In Sec.XI we present our conclusions.

\section{ Isolated magnetoexciton in the system of spatially separated
electrons and holes}\label{Sin_Ex}

The total Hamiltonian $\hat H$ of an isolated pair of spatially
separated $e$ and $h$ in the magnetic field is:

\begin{equation}\label{H_Main}
\hat H=
\frac{1}{2m_{e}}\left( -i \nabla_e + e{\bf A_e} \right)^2 +
\frac{1}{2m_{h}}\left( -i \nabla_h - e{\bf A_h} \right)^2 -
\frac{e^2}{\sqrt{({\bf r_e - r_h})^{2} + D^{2}}},
\end{equation}

where  $m_{e}$, $m_{h}$ are the effective electron and hole masses; ${\bf
A_e}$, ${\bf A_h}$ are the vector potentials in electron and hole
location, respectively;
${\bf r_e}$, ${\bf r_h}$ are electron and hole locations
along quantum wells (we use units $c = \hbar = 1$).

A conserved quantity for isolated exciton in magnetic field
(the exciton {\it magnetic} momentum)
 is (see Ref.[\onlinecite{GorDzyal}]):

\begin{equation}\label{Momentum}
\hat {\bf P} = -i \nabla _{e} -i \nabla _{h} + e ({\bf A_{e}} -
{\bf A_{h}}) - e[{\bf H, r_{e} - r_{h}}]  ,
\end{equation}

The conservation of this quantity is  connected with the
invariance of the system upon a simultaneous translation of $e$ and $h$
and gauge transformation.

Let us consider the coordinates of the center of mass ${\bf R} =
\frac{m_{e}{\bf r_e} + m_{h}{\bf r_h}}{m_{e} + m_{h}}$
 and the internal exciton coordinates
 ${\bf r} = {\bf r_e} - {\bf r_h}$.
The cylindrical gauge for vector-potential is used:  ${\bf A_{e,h}} =
\frac{1}{2} [{\bf H,r_{e,h}}]$.

Eigenfunctions of Hamiltonian Eq.(\ref{H_Main}), which are also the
eigenfunctions of the magnetic momentum $\bf{P}$),
are (see \cite{Lerner}$^,$\cite{GorDzyal}):

\begin{equation}\label{W_Func_Gen}
\Psi _{Pk} ({\bf R,r}) = \exp \left\{i{\bf R} \left( {\bf P} +
\frac{e}{2} [{\bf H},{\bf r}] \right)  + i \gamma \frac{{\bf Pr}}{2} \right\}
\Phi _k ({\bf P,r}) ,
\end{equation}

where $\Phi _k ({\bf P,r})$ is the function of internal coordinates
${\bf r}$; ${\bf P}$ is the eigenvalue  of magnetic momentum; $k$ are
quantum numbers of exciton internal motion. In high magnetic fields
 $k = (n,m)$, where  $n = min(n_{1}, n_{2})$, $m = |n_{1} - n_{2}|$, $n_{1,
2}$ are Landau quantum numbers for $e$ and $h$
\cite{Lerner}$^,$\cite{Ruvinskiy}; $\gamma = \frac{m_h - m_e}{m_h + m_e}$.

The effective Hamiltonian  $H_{{\bf P}}$ has the form:

\begin{equation}\label{H_P}
\begin{array}{c}
\hat H_P =  -\frac{1}{2\mu ^{*}} \Delta _r -
\frac{ie \gamma }{2\mu ^{*}} [{\bf H,r}] \nabla _r
+ \frac{e^2}{8\mu ^{*}} [{\bf H,r}]^2 +
\frac{e}{M} [{\bf H,r}]{\bf P}
\\   \\
+\frac{{\bf P}^2}{2M}
-\frac{e^2}{\sqrt{{\bf r}^2 + {\bf D}^2}} ,
\end{array}
\end{equation}

where $\mu ^{*} = \frac{m_{e}m_{h}}{m_{e} + m_{h}}$.

 Using the Feynman theorem
(we denote $\langle Pk | ... | Pk \rangle $ as $\langle  ... \rangle $)
one can obtain for isolated magnetoexciton current
 (see Ref.[\onlinecite{GorDzyal}]):

\begin{eqnarray}
\label{cur}
{\bf J _k(P)} =
M \langle  \frac{\partial \hat H_P}{
\partial {\bf P}}   \rangle = M
\frac{ \partial \varepsilon _k (P)}{\partial {\bf P} }
= M \frac{{\bf P}}{P}
\frac{\partial  \varepsilon _{k}(P)}{\partial P} ,
\end{eqnarray}

where $M = m_e + m_h$; $\varepsilon _{k}(P)$ is the magnetoexciton
dispersion law (for indirect excitons $\varepsilon _{k}(P)$ in dependence on
$H$ and interwell separations $D$ was analyzed in detail \cite{Ruvinskiy}).

The dispersion relation  $\varepsilon _{k}(P)$  of isolated magnetoexciton
is the quadratic function at small magnetic momenta:

\begin{equation}\label{Energy}
\varepsilon _{k}({\bf P}) \approx \frac{P^2}{2m_{H k}},
\end{equation}

where $m_{H k}$ is the effective {\it magnetic} mass, dependent  on $H$
and the distance $D$ between $e$ -- and $h$ -- layers and quantum number
$k$ (see Ref.[\onlinecite{Ruvinskiy}]).

The quadratic dispersion relation Eq.(\ref{Energy}) is true for small $P$
at arbitrary magnetic fields $H$ and follows from the fact that $P = 0$
is an extremal point of the  dispersion law $\varepsilon _{k}({\bf P})$.
The last statement may be proved by taking into account the regularity
of the effective Hamiltonian  $H_{{\bf P}}$
 as a function of the parameter ${\bf P}$ at ${\bf P} = 0$ and also the
invariance of $H_{{\bf P}}$ upon simultaneous rotation of ${\bf r}$
and ${\bf P}$  in the CQW plane.  \cite{yel}
 For magnetoexciton ground state $m_{H} > 0$.

For high magnetic fields $r_{H} \ll a_{0}^{*}$  and at $D \lesssim r_{H}$
the quadratic dispersion relation is valid at $P \ll \frac{1}{r_{H}}$,
but for $D \gg r_{H}$ it holds over a wider region --- at least at $P \ll
\frac{1}{r_{H}} \frac{D}{r_{H}}$ \cite{Ruvinskiy} ($a_{0}^{*} =
\frac{1}{2\mu e^{2}}$ is the radius of a 2D exciton at $H = 0$; $\mu =
\frac{m_{e}m_{h}}{m_{e} + m_{h}}$; $m_{e,h} $ are the effective masses of
$e$ and $h$).

Using the quadratic dispersion relation for magnetoexcitons, one has at
any $H$  an expression for the magnetoexciton velocity  analogous to
that for the ordinary momentum ${\bf \dot R} = \frac{\partial \varepsilon
_{k}({\bf P})}{\partial {\bf P}} = \frac{{\bf P}}{m_{H k}}$. So the mass
current of an isolated magnetoexciton for small $P$ is:

\begin{equation}\label{J_P}
{\bf J}_{k} ({\bf P}) = \frac{M}{m_{H k}} {\bf P} .
\end{equation}

\section{Spectrum of collective excitations}

Due to interlayer separation  $D$ indirect magnetoexcitons
both in ground state and in excited states have electrical dipole moments.
We suppose, that indirect excitons interact as {\it parallel} dipoles.
 This is valid,
when $D$ is larger than the mean separation $\langle  r \rangle $
between electron and hole
along quantum wells $D \gg \langle  r \rangle$. We take into
account that at high magnetic fields $\langle  r \rangle \approx
Pr_H^2 $ ($\langle {\bf r} \rangle $ is normal to ${\bf P}$)
 and that the typical value of magnetic
momenta (with exactness to logarithm of the exciton density
($ln(n_{ex})$, see below) is $P \sim \sqrt{n_{ex}}$ (if the dispersion
relation $\varepsilon _{k} (P) = \frac{P^2}{2m_{H k}}$ is true).  So the
inequality $D \gg \langle r \rangle$ is valid at $D \gg \sqrt{n}
r_H^2 $.

The distinction between excitons and bosons manifests itself in
exchange effects
(see, e.g., \cite{KelKoz}$^,$\cite{Berman}).  These effects for excitons
with spatially separated $e$ and $h$ in a rare system $n_{ex}a^{2}(H,D) \ll 1$
are suppressed due to the negligible overlapping of wave functions of two
excitons on account of the potential barrier, associated with the
dipole-dipole repulsion of indirect excitons \cite{Berman}
(here $n_{ex}$, $a(D,H) $ are respectively
 density and magnetoexciton radius along quantum
wells, respectively). Small
tunnelling parameter connected with this barrier is $exp[-
\frac{1}{\hbar }\int_{a(H,D)}^{r_{0}}\sqrt{2m_{H
k}(\frac{e^{2}D^{2}}{R^{3}}
- \frac{\kappa ^{2}}{2m_{H k}})}
dR]$, where $\kappa ^{2} \sim \frac{n}{\ln \left(
\frac{1}{8\pi n m_{H k}^2 e^4 D^4} \right)}$ is the characteristic
momentum of the system (see below); $r_{0} = (2m_{H k}e^{2}D^{2}/\kappa
^{2})^{1/3}$ is the classical turning point for the dipole-dipole
interaction.  In high magnetic fields the small parameter mentioned above
has the form
$exp[-2\hbar ^{-1}(m_{H k})^{1/2}eD a^{-1/2}(H,D)]$. Therefore the
system of indirect magnetoexcitons can be treated by the diagram
technique for a boson system.

But in contrast with a 2D boson  system in the absence of magnetic field
(see Ref.[\onlinecite{Ydson}]),
 some problems arise due to nonseparation of the
relative motion of e and h and exciton center of mass motion in magnetic
fields.  \cite{Lerner}$^,$\cite{Ruvinskiy}
  Due to the nonseparation of internal and center of
mass motions  the Green functions depend on both
the external coordinates ${\bf R}, {\bf R}'$ and the internal coordinates
${\bf r}, {\bf r}'$.

For the rare two-dimensional
 magnetoexciton system (at $n_{ex} a^{2}(D,H) \ll 1$)
 the summation of ladder diagrams is  adequate.
 The integral
equation for vertex $\Gamma $ in the ladder approximation is represented
 on Fig.1. In the strong magnetic fields the representation using as a
basis of isolated magnetoexciton wave functions $\Psi _{{\bf P},m}({\bf r},
{\bf R})$ is convenient.

We use the following approximation for the interaction between two
 magnetoexcitons $U(P) = U_{0}$ at $P < a^{-1}(H,D)$ and
$U(P) = 0$ at $P > a^{-1}(H,D)$.
After exciton-exciton scattering their total magnetic momentum is
conserved, but  magnetic momentum of each exciton can be changed.  Since the
mean distance between $e$ and $h$ along quantum wells is proportial to
the magnetic momentum, the scattering is accompanied by the exciton
polarization.  We consider sufficiently low temperatures when
magnetoexciton states with only small magnetic  momenta $P \ll
\frac{1}{r_{H}}$ are filled. The change of these magnetic momenta due to
exciton-exciton scattering  is also negligible due to the conservation of
the total magnetic momentum. But these small magnetic momenta correspond
to small separation between electrons and holes along quantum wells $\rho
\ll r_{H}$.  So magnetoexciton polarization due to scattering is
negligible and the magnetoexciton dipole moment keeps to be almost normal to
quantum wells $d = eD$, i.e. the interexciton
interaction law is not changed due to the
scattering.

The equation for $\Gamma $ can be solved in the strong magnetic fields
 $\omega _c = eH/\mu ^{*} \gg e^2/r_H$,
when the characteristic value  of $e-h$ separation in
the magnetoexciton $| \langle {\bf r} \rangle |$ has the order of the
magnetic length $r_H = 1/\sqrt{eH}$. The functions
$\Phi _k ({\bf P,r}) $ (see Eq.(\ref{W_Func_Gen}))
 are dependent on the difference $({\bf r - \rho})$,
where ${\bf \rho} = \frac{r_H^2}{H} [{\bf H,P}]$ \cite{Ler},
\cite{GorDzyal}.  At small magnetic momenta $P \ll 1/r_H$ we have: $\rho
\ll r_H$, and, therefore, in functions $\Phi _k ({\bf r-\rho}) $ we can
ignore the variable ${\bf \rho}$ relatively to ${\bf r}$. In the strong
magnetic field quantum numbers $k$ correspond to the quantum numbers
$(m,n)$ (see above). For the lowest Landau level we denote
$\varepsilon _{00}({\bf P}) = \varepsilon ({\bf P})$ and ${\bf J}_{00}({\bf
P}) = {\bf J}({\bf P})$.  Using the orthonormality of functions $\Phi
_{mn} ({\bf 0,r})$ we obtain  approximate
equation for the vertex $\Gamma $ in
strong magnetic fields.
In high magnetic field, when the typical interexciton interaction
$D^{2}n^{-\frac{3}{2}} \ll  \omega _{c}$,
 one can ignore transitions between Landau levels and consider
only the states on the lowest Landau level $m=n=0$.  Since typical value
of $r$ is $r_H$, and $P \ll 1/r_H$ in this approximation the equation
for the vertex in the magnetic momentum representation  $P$ (see Fig.1.)
on the lowest Landau level $m=n=0$
has the same form  (compare with Ref.[\onlinecite{Phys}])
 as for two-dimensional boson
system without a magnetic field, but with  the magnetoexciton magnetic
mass $m_{H}$ (which depends on $H$ and $D$) instead of the exciton mass
($Π= m_{e} + m_{h}$) and magnetic momenta instead of ordinary momenta:

\begin{equation}\label{Gamma_Int}
\begin{array}{c}
\Gamma (k,q;L ) = U _F ({\bf k - q })  +
\int_{}^{} \frac{d {\bf l}}{(2\pi )^2}
\frac{U_F ({\bf k-l}) \Gamma (l,q;L )}{\frac{\kappa ^2}{m_H}
+\Omega - \frac{L^2}{4m_H} - \frac{l^2}{m_H} +i\delta }
\\   \\
\mu = \frac{\kappa ^2}{2m_{H}} = n_{ex}\Gamma _{0}.
\end{array}
\end{equation}
Here $\mu $ is the chemical potential of the system.

We find the solution of Eq.(\ref{Gamma_Int}) by using the approximation
for the effective interaction:

\begin{equation}\label{approx}
\Gamma (P) = \left\{
\begin{array}{rl}
\Gamma _{0}, & P < a^{-1}(H,D), \\
          0, & P > a^{-1}(H,D).
\end{array}  \right.
\end{equation}

The integral equation Eq.(\ref{Gamma_Int}) for the vertex  can be solved
analytically in the approximation  $\kappa \ll \sqrt{n}$. This inequality
must be fulfilled  simultaneously with the condition of low density
$na^{2}(H,D) \ll 1$ which is necessary for the applicability of
the ladder approximation.  The
solution of the integral equation for the vertex $\Gamma $ of this system
can be  expressed through the solution of the equation for the amplitude
of scattering $f_0(\kappa ) $ of isolated pair of interacting
particles (with a mass equal to the magnetic mass $m_{H}$ of magnetoexciton)
 in the  two-dimensional system {\it without magnetic field}
 with the repulsing
potential $U(R) = e^2 D^2/R^3$:

\begin{equation}\label{f_0}
f_0 (\kappa) = \frac{\left( \frac{\pi i}{2 \kappa } \right) ^{1/2}}
{\ln (\kappa m_H e^2 D^2)} .
\end{equation}

Here the characteristic magnetic momentum $\kappa $
contrary to three-dimensional system is not equal to zero and is determined
from the relation:

\begin{equation}\label{Kappa}
\kappa ^2 = -4nf_0 (\kappa) \left(\frac{2\pi \kappa }{i} \right)^{1/2} .
\end{equation}

This is specific feature of two-dimensional
Bose system connected with logarithmic divergence of two-dimensional
scattering amplitude at zero energy.
A simple analytical solution for the chemical potential can be obtained
if $\kappa m_H e^{2} D^2 \ll 1$.  In strong magnetic fields at $D \gg
r_{H}$ the exciton magnetic mass is $m_H \approx
\frac{D^{3}}{e^{2}r_{H}^{4}}$ \cite{Ruvinskiy}. So the inequality $\kappa
m_{H}e^{2}D^{2} \ll 1$ is valid if $D \ll (r_{H}^{4}/n^{1/2})^{1/5}$. In
result the chemical potential $\mu $ is obtained in the form:

\begin{equation}\label{Mu}
\mu =  \frac{\kappa ^2 }{2m_{H}}
= \frac{8\pi n}{2m_H \ln \left( \frac{1}{8\pi n m_H^2 e^4 D^4} \right)} .
\end{equation}

The spectrum of collective excitations following from the solution of
Eq.(\ref{Gamma_Int}) at small magnetic momenta is the sound $\varepsilon
(P)  = c_s P$ with the sound velocity $c_s = \sqrt{\frac{n\Gamma }{m_H}}
= \sqrt{\frac{\mu}{m_H}}$, where $\mu $ is defined by Eq.(\ref{Mu}).

\section{The density of the superfluid component}

The temperature of the Kosterlitz-Thouless transition
$T_c $ \cite{Kosterlitz} to the superfluid state
in a two-dimensional magnetoexciton system is determined by the
equation:

\begin{equation}\label{T_KT}
T_c = \frac{\pi \hbar ^2 n_s (T_c)}{2 k_B m_{H}},
\end{equation}

where $n_s (T)$ is the
superfluid density of the magnetoexciton system  as a function
of  temperature $T$, magnetic field $H$ and interlayer distance
$D$; and $k_B$ is Boltzmann constant.

The function $n_s (T)$ (\ref{T_KT}) can be found from
the relation
$n_s = n_{ex} - n_n $ ($n_{ex}$ is the total density, $n_{n}$ is
the normal component density).
We determine the normal component density by the usial procedure
\cite{Abrikosov}. Suppose that the magnetoexciton system  moves with
a velocity $u$. At nonzero temperatures dissipating quasiparticles
will appear in this system. Since their density is small at low
temperatures, one can assume that the gas of quasiparticles is an ideal
Bose gas.

To calculate the superfluid component density
we find the total current of quasiparticles in a
 frame in which the superfluid component is at rest.
Then using Eq.(\ref{cur})  we obtain
the mean total current of 2D magnetoexcitons in the coordinate system,
moving with a velocity ${\bf u}$:

\begin{eqnarray}
\label{nnor}
\begin{array}{c}
\langle {\bf J}' \rangle = \frac{M}{m_H} \langle {\bf P}' \rangle =
\int_{}^{} \frac{d{\bf P'}}{(2\pi ) ^{2})} {\bf P}'
f_{\varepsilon (P') - {\bf P'u} }.
\end{array}
\end{eqnarray}

In the first order by ${\bf P'u}/T$ we have:

\begin{equation}\label{J_Tot}
\langle {\bf J}' \rangle =
{\bf u} \frac{M}{m_H} \frac{3 \zeta (3) }{2 \pi } \frac{T^3}{c_s^4}
\end{equation}

Using Eq.(\ref{J_Tot}), we see that the total current of the system
is proportional to the total magnetic
momentum ${\bf P}$. Then we determine the
normal component density $n_{n}$:

\begin{equation}\label{J_M}
\langle {\bf J}' \rangle = M n_n {\bf u} .
\end{equation}

Comparing Eqs.(\ref{J_M}) and (\ref{J_Tot})  we obtain the expression
for the normal density $n_{n}$.
As a result,  we have for the superfluid density:

\begin{equation}
\label{n_s}
n_s = n_{ex} - n_n =
n_{ex} - \frac{3 \zeta (3) }{2 \pi } \frac{T^3}{c_s^4 m_H}.
\end{equation}

It occurs that
the expression for the superfluid density $n_{s}$
in the strong magnetic field for the magnetoexciton
rare system differs from analogous
expression in the absence of  magnetic field (compare
with Ref.[\onlinecite{Berman}]) by replacing of exciton mass $M = m_{e} +
m_{h}$ with the exciton magnetic mass $m_{H}$.

\section{ Superfluid state transition}

{\bf a. Kosterlitz-Thouless temperature}

The superfluidity of magnetoexcitons appears below the
Kosterlitz-Thouless temperature $T_{c}$ (Eq.(\ref{T_KT})),
where only bound vortexes are present.
Substituting the expression for the superfluid component density
$n_{s}$ from Eq.(\ref{n_s}) into Eq.(\ref{T_KT}),  we obtain
the equation for the Kosterlitz-Thouless temperature $T_c$.
The solution is:

\begin{equation}
\label{tct}
\begin{array}{c}
T_c =
[\left( 1 + \sqrt{\frac{16}{(6\cdot 0.45)^{3}\pi ^{4}}
(\frac{m_{H}T_{c}^{0}}{n_{ex}})^{3} +
1} \right)^{1/3}   +
\left( 1 -  \sqrt{\frac{16}{(6\cdot 0.45)^{3}\pi ^{4}}
(\frac{m_{H}T_{c}^{0}}{n_{ex}})^{3} +  1} \right)^{1/3}] \frac{T_{c}^{0}}{
(4\pi )^{1/3}}   .
\end{array}
\end{equation}

Here $T_{c}^{0}$ is the auxiliary quantity equal to the temperature  of
vanishing of  superfluid density in the mean field approximation
 $n_{s}(T_{c}^{0}) = 0$:

\begin{equation}
\label{tct0}
T_c^0 = \left( \frac{2 \pi n_{ex} c_s^4 m_H}{3 \zeta (3)} \right)^{1/3} =
\left( \frac{32  }{3 \zeta (3)
ln^{2}(\frac{1}{8\pi n m_{H}^{2}D^{4}})} \right)^{1/3}
\frac{\pi n_{ex}}{m_{H}}  .
\end{equation}

In high fields  $H$ and at small ${\bf P}$
for the lowest Landau level  ($n = 0$) and at quantum number $m = 0$
the exciton effective magnetic mass is
$ m_{H} = \frac{2^{3/2}}{e^{2}r_{H}\sqrt{\pi }}$ at $D \ll r_{H}$
and $ m_{H} \approx \frac{D^{3}}{e^{2}r_{H}^{4}}$ at $D \gg r_{H}$.
At large $D$, i.e., for $D \gg a_{0}^{*}$ in weak
 fields ($r_{H} \gg a_{0}^{*}$)  or  $D \gg r_{H}$ in
high fields ($r_{H} \ll a_{0}^{*}$) one has $m_{H} =
M + \frac{H^{2}D^{3}}{c^{2}} $. \cite{Ruvinskiy}

The temperature $T_{c}^{0} = T_{c}^{0}(D,H)$ may be used as a crude estimate
of the crossover region where local superfluid density appeares
for rare magnetoexciton system on the scales smaller or
of order of mean intervortex
separation in the system. The local superfluid density can manifest itself
in local optical properties or local transport properties (see below).
In rare two-dimensional system in the ladder approximation
(i.e. at $\ln (8\pi n_{ex} m_H^2 e^4 D^4)^{-1} \gg 1 $)
 the Kosterlitz-Thouless temperature is:

\begin{eqnarray}
\label{strong}
T_{c} = \frac{T_{c}^{0}}{(2\pi )^{1/3}}  .
\end{eqnarray}

  At maximal temperature of superfluid
(the Kosterlitz-Thouless temperature) the normal density approximately
is

\begin{eqnarray}
\label{quasip}
n_{n}(T_{c}) = \frac{3 \zeta (3) }{2 \pi } \frac{T_{c}^{3}}{c_s^4 m_H}  .
\end{eqnarray}

This estimate does not take into account small contribution of vortexes.
Substituting Eq.(\ref{tct}) to Eq.(\ref{quasip}), we obtain

\begin{equation}
\label{quasipa}
\begin{array}{c}
n_{n}(T_{c}) = \frac{n_{ex} }{4 \pi }
[\left( 1 + \sqrt{\frac{16}{(6\cdot 0.45)^{3}\pi ^{4}}
(\frac{m_{H}T_{c}^{0}}{n_{ex}})^{3} +
1} \right)^{1/3}   +           \\
\left( 1 -  \sqrt{\frac{16}{(6\cdot 0.45)^{3}\pi ^{4}}
(\frac{m_{H}T_{c}^{0}}{n_{ex}})^{3} +  1} \right)^{1/3}]^{3}   .
\end{array}
\end{equation}

In \, rare \, two-dimensional \, system \, in \, the \, ladder \,
approximation \,  we have:

\begin{eqnarray}
\label{quasipar}
\frac{n_{n}(T_{c})}{n_{ex}} = \frac{1 }{2 \pi }   .
\end{eqnarray}

Note that Eqs.(\ref{strong}) and (\ref{quasipar}) take place
for {\it any} two-dimensional rare Bose gas.
The dimensionless value $n_{n}(T_{c})/n_{ex}$ can be considered as the
small parameter.
So the approximation of the ideal Bose gas of quasiparticles
is valid.
Note that for the dense electron-hole system without magnetic field at
$n_{ex} \to \infty$ opposite case $\frac{n_{n}(T_{c})}{n_{ex}} \to 1$
takes place
due to exponential vanishing of the order parameter $\Delta $
 (see, e.g., Refs.[\onlinecite{Ydson}] and
[\onlinecite{Berman}]).

According to Eqs.(\ref{tct}) and (\ref{tct0}) the temperature
of the onset of superfluidity due to the Kosterlitz-Thouless
transition at {\it a fixed magnetoexciton density}
decreases as a function of magnetic field
due to the increase in $m_{H}$ as a function of $H$ and $D$, while $T_{c}$
decreases
as $H^{-{1}/{2}}$ at $D \ll r_{H}$ or as $H^{-2}$ at $D \gg
r_{H}$, and $n_{s}$ is a slowly decreasing function of $D$.
The dependencies of  $T_{c}$ on  $H$ are shown in Fig.2.

From Eqs. (\ref{tct}) and (\ref{tct0}) one can see that
the Kosterlitz-Thouless temperature of a rare magnetoexciton system
is proportional to the
magnetoexciton density $n_{ex}$. At  high magnetic fields
the symmetry $\nu \to 1 - \nu $,  $e \longleftrightarrow h$ takes place for
 the Landau level (see Ref.[\onlinecite{Lerner}]). Thus unoccupied states on
Landau levels for spatially separated electrons and holes can bind
to "antiexcitons" and superfluidity of "antiexcitons"  can also take place
at $1 - \nu \ll 1$. The Kosterlitz-Thouless temperature  for
superfluidity of
 antiexcitons as function of
 $H$, $D$ for strong $H$ is symmetrical to that for excitons.
The top Kosterlitz-Thouless
 temperature at high magnetic fields corresponds to the "maximal"
density $n_{max}$ of stable magnetoexciton system at the Landau level
$n_{max} = \nu _{max} \frac{1}{4\pi r_{H}^{2}} \sim H$, where
$\nu _{max}(D)$ is the maximal filling of Landau level for magnetoexcitons
- see below
(for "antiexcitons" the corresponding critical value is $1 - \nu _{max}(D)$).

{\bf b. The problem of large magnetic momenta}

At large magnetic momenta $P$ the isolated magnetoexciton spectrum
$\varepsilon (P)$ contrary to the case $H = 0$ has a constant limit
(being equal to Landau level $\frac{\hbar \omega _{c}}{2}$
for reduced effective mass, see Refs.[\onlinecite{Ruvinskiy}],
[\onlinecite{Lerner}]).  As
 a result the spectrum of {\it interacting} magnetoexciton
system also have a plateau at great momenta. So Landau criterium of
superfluidity is not valid at large $P$ for the interacting magnetoexciton
system. However the probability of excitation of quasiparticles  with
magnetic momenta $P \gg 1/r_{H}$ by moving magnetoexciton system is
negligibly small at small superfluid velocities. In this sense, the
superfluidity of 2D magnetoexcitons keeps to be almost metastable one.
This can be shown by the estimation of the probability $d W$ of
the excitation  of the quasiparticle on the plateau with magnetic momenta
$P \gg 1/r_{H}$; the energy of quasiparticles on the plateau
$\varepsilon (P)$ equals to the magnetoexciton binding energy.
At high magnetic fields we  have:

\begin{equation}\label{e_quasi}
\begin{array}{c}
\varepsilon (P) \sim \sqrt{\frac{\pi }{2}}\frac{e^{2}}{r_{H}}
- \frac{e^2}{Pr_H^2},
\quad D \ll a(H,D), \quad P r_{H}^{2} \gg D
\\  \\
\varepsilon (P) \sim \frac{e^{2}}{D}
- \frac{e^2}{Pr_H^2}, \quad D \gg a(H,D), \quad P r_{H}^{2} \gg D .
\end{array}
\end{equation}

At the motion of magnetoexciton liquid in a lattice with the small velocity
${\bf u}$, which is smaller than the sound velocity $c_{s}$, according to
the Landau criterium \cite{Abrikosov}  creation of the quasiparticles in
the region of plateau at great momenta with the magnetic momentum
$P \gg 1/r_{H}$ and the energy $\sim E_{0}$ is possible due to the
friction  between liquid and impurities, defects in  the lattice or
roughness of boundaries of quantum wells. So when one quasiparticle
appears the liquid gets the magnetic momentum $P$. The appearance of
the large magnetic momentum in the liquid  is equivalent to the great
mean separation between electron and hole along one layer \\ ${\bf \rho } =
\frac{r_{H}^{2}}{H} [${\bf H}, ${\bf P}]$ (see Sec.II)
 So magnetoexcitons with {\it very} large $P$ does not exist due to the
interaction of electron and hole with impurities etc.

Let us estimate the probability $d W_{P}$ of the transition of the
superfluid system from the initial state  with the magnetic momentum
$P = 0$
without quasiparticles to the final state  with one quasiparticle  with the
large magnetic momentum $P \gg 1/r_{H}$ by using
Fermi golden rule taking into account
the "friction" interaction $V$. We have for the probability per unit
of time $d W(P)$:

\begin{equation}\label{fermi}
d W(P) = \frac{2\pi}{\hbar} | \langle 0 | \hat V \hat \alpha ^{\dag}
| 0 \rangle   | ^2
\delta (  \Delta E_{k} + \varepsilon (P) + {\bf Pu} )
d \nu _{\varepsilon } ,
\end{equation}
where $\nu _{\varepsilon }$ is the density of final states of the system;
$\Delta E_{k}$ is the change in the kinetic
energy of superfluid liquid; $\hat V $ is the
"friction" interaction (see below);
 $|0\rangle $ is a
ground state of magnetoexciton superfluid;
$\alpha _{\bf P} ^{\dag}$ is the quasiparticle creation operator.
After quasiparticle creation total magnetic momentum of the system
is conserved.

 At large momentum $P \gg 1/r_H$ the wave
function of quasiparticle is almost the same as wave function of the isolated
magnetoexciton.
 It means that the quasiparticle annihilation
operator $\alpha _{\bf P}$ is almost the
 same as the ordinary particle annihilation
operator $a _{\bf P}$.

In second quantified representation the "friction" interaction operator
$\hat V$ can be represented as

\begin{equation}\label{Friction}
\hat V = \sum \limits ^{} _{\bf P'P''}
V _{\bf P'P''} a ^{\dag } _{\bf P'} a _{\bf P''}
\end{equation}
where $V _{\bf P'P''} $ is the matrix element of "friction" interaction
calculated with the use of isolated magnetoexciton eigenfunctions
Eq.(\ref{W_Func_Gen}).

We find:

\begin{equation}\label{Matrix_2}
\langle 0 | \hat V a ^{\dag} _P | 0 \rangle =
V _F ({\bf P}) e ^{-\frac{1-2 \gamma }{8} P^2 r_H^2} ,
\end{equation}

where $V _F ({\bf P})$ is the Fourier-transform of $V$.
Then the probability $d W_{P}$ of the creation of the quasiparticle
per unit of time with the large magnetic momentum $P$ and the
energy $\varepsilon (P)$ is:

\begin{eqnarray}\label{probfinis}
d W(P) = \frac{1}{(2 \pi )^2 \hbar ^3}
e^{-\frac{1+2\gamma ^2}{4}r_{H}^{2} P^{2}}
| V_F ({\bf 0,P}) | ^2
\delta ( \Delta E + \varepsilon (P) + {\bf Pu} ) P dP,
\end{eqnarray}
Thus the probability $d W_{P}$  of the creation of the
quasiparticle with the large magnetic momenta $P \gg 1/r_{H}$
 is negligibly small as
$d W_{P} \sim e^{-\frac{1+2\gamma ^2}{4}r_{H}^{2} P^{2}} \ll 1$. So the
superfluidity of 2D magnetoexcitons keeps to be almost metastable one.
Note that at small magnetic momenta $P \ll 1/r_{H}$ in the region of
the sound spectrum of interacting magnetoexcitons
Landau criterium of
superfluidity is valid and the
probability $d W_{P}$  of the creation of the
quasiparticle in the region of the sound spectrum at $u < c_{s}$ is zero
due to $\delta (  \Delta E + \varepsilon (P) + {\bf Pu} ) = 0$
in Eq.(\ref{fermi}).

At low temperatures $T < T_{c} \ll E_{0}$
states with large magnetic momenta are negligibly filled
($exp[- \frac{\varepsilon (P)}{T}] \ll 1$, where
$\varepsilon (P) $ is the magnetoexciton energy which has
the same order as magnetoexciton binding energy $E_0$; at high
magnetic fields $E_{0} = \sqrt{\frac{\pi }{2}}\frac{e^{2}}{r_{H}}$
at $D \ll a(H,D)$ and $E_{0} = \frac{e^{2}}{D}$
at $D \gg a(H,D)$). So quasiparticles at large magnetic momenta $P$
give a small contribution to the  densities  of the normal component $n_{n}$
and superfluid component $n_{s}$ (see Eq.(\ref{n_s})). Hence,
the expressions given above
for the temperature of Kosterlitz-Thouless transition
are valid.

\section{Thermodynamics and equation of the state of the system
at high temperatures}

Now we estimate correction terms in the
chemical potential and the
equation of the state for slightly nonideal gas of 2D magnetoexcitons
at high temperatures due to exchange effects and dipole-dipole interaction
between magnetoexcitons. We show these correction terms are small
at high magnetic fields and high temperatures,
and so their contributions
to the chemical potential and the equation of the state   are
additive. So we can consider these effects separately.

One has for the free energy $F$ of ideal gas of bosons:  \cite{Landaul}

\begin{eqnarray}
\label{fr}
F = F_{Bol}(1 - \frac{1}{4}e^{\mu /T})  .
\end{eqnarray}

The chemical potential $\mu ^{0}$ of ideal gas of magnetoexcitons
can be obtained
from the normalization condition for the number of magnetoexcitons:

\begin{eqnarray}
\label{che}
\mu ^{0} =  - T ln(\frac{m_{H}T}{2\pi \hbar ^{2} n_{ex}})   .
\end{eqnarray}

At high temperatures and high magnetic fields
the inequality $e^{\mu /T} \ll 1$ is true for a rare system.
Using the relation for the pressure
$P = - (\frac{\partial F}{\partial S})_{T,N}$ and Eq.(\ref{fr}),
 we have for the equation
of the state

\begin{eqnarray}
\label{eqb}
P = \frac{N T}{S} (1 - \frac{\pi \hbar ^{2} n_{ex}}{2m_{H}T}).
\end{eqnarray}

Using the relation $\mu = (\frac{\partial F}{\partial N})_{T,S}$
for the chemical potential $\mu $, we obtain the contribution
of exchange interactions to the chemical potential $\mu $
(with exactness to $O(n^{2})$):

\begin{eqnarray}
\label{mexdip}
\mu =  - T ln(\frac{m_{H}T}{2\pi \hbar ^{2} n_{ex}})
+ 2 T  \frac{\pi \hbar ^{2} n_{ex}}{2m_{H}T} .
\end{eqnarray}

Now  we analyze the contribution of interaction. We estimate
 for rare 2D magnetoexciton system
 the second virial coefficient $B(T)$ in expansion of 2D pressure
 on $1/S$ ($S$ is the area of the system;
Boltzmann constant $k_{B} = 1$):

\begin{eqnarray}
\label{pr}
P = \frac{N T}{S} (1 + \frac{N B(T)}{S} + ... ).
\end{eqnarray}

At high temperatures the virial coefficient is:

\begin{eqnarray}
\label{vir}
B(T) = \frac{1}{2} \int_{}^{} (1 - e^{-U(R)/T}) dS
 \approx \pi T^{-2/3}(eD)^{4/3} ln(\frac{2\pi \hbar ^{2} n_{ex}}{m_{H}T}) ,
\end{eqnarray}

where $U(R) = \frac{e^{2}D^{2}}{R^{3}}$ is the pair interaction
 between particles. We integrate Eq.(\ref{vir}) by coordinate
from the classical turning point for the dipole-dipole interaction
 $R_{0} = (e^{2}D^{2}/\mu )^{1/3}$, substituting the chemical potential
$\mu $  Eq.(\ref{che}).
At high temperatures
 $U(R_{0})/T \ll 1$ (where $R_{0} \sim (\pi n)^{-1/2}$).

Using  additivity of small
exchange  and dipole-dipole interaction corrections, we
have the equation of the state with both corrections included:

\begin{eqnarray}
\label{eqbexch}
P S = N T (1 - \frac{\pi \hbar ^{2} n_{ex}}{2m_{H}T} +
\pi T^{-2/3}(eD)^{4/3} n_{ex} ln(\frac{2\pi \hbar ^{2} n_{ex}}{m_{H}T})).
\end{eqnarray}

For the chemical potential $\mu $, we obtain
with exactness to $O(n^{2})$
the chemical potential with both terms included:

\begin{eqnarray}
\label{ch}
\mu =  - T ln(\frac{m_{H}T}{2\pi \hbar ^{2} n_{ex}})
+ 2 \pi T^{1/3}(eD)^{4/3} n_{ex} ln(\frac{2\pi \hbar ^{2} n_{ex}}{m_{H}T})
+ 2T \frac{\pi \hbar ^{2} n_{ex}}{2m_{H}T}.
\end{eqnarray}

  The virial coefficient
$B(T)$ decreases {\it vs.} $H$ due to increase of magnetic mass $m_{H}$.
Hence, in high magnetic fields the system of indirect magnetoexcitons
is almost ideal gas due to $B(T) \ll 1$.
Decrease of interexciton interaction can be investigated
experimentally. In noninteracting system
shape of spectral lines of magnetoexciton photoluminescence is determined by
Doppler effect. For an interacting system
  it was demonstrated \cite{Weiss} that the
shape of lines of photoluminescence of
 semiconductor quantum wells
is dominated by many-body interactions, and it is
 essentially different from isolated particles because of
exciton-exciton interactions.
  Comparing line shapes
at the different magnetic fields, one can demonstrate transition to an
 ideal gas in the exciton system in high magnetic fields.

\section{Region of existence of magnetoexciton phase}

{\bf a. Ionization of magnetoexcitons  in the quasiclassical region}

In magnetoexciton system in coupled  quantum wells at essentially higher
temperatures than the
transition to the superfluid state the
exciton thermal ionization takes place.
Magnetoexcitons existence line $T_{i}(H,n)$ (more strictly, crossover
region) can be obtained in quasiclassical regime from the
ionization equilibrium condition analogous to Saha relation \cite{Landaul}
from condition
of equality of exciton chemical potential to the sum
of chemical potentials of electrons and holes.

We neglect transitions between Landau levels in high magnetic fields
$\hbar \omega _{c} \gg T$ ($\omega _{c} = \frac{eH}{m_{e}c}$ is the cyclotron
energy) and are measured the chemical potentials of electrons $\mu _{e}$,
holes $\mu _{h}$ and magnetoexcitons $\mu _{ex}$ from the
lowest Landau level. Then we have for the chemical potentials of electrons
and holes at $2\pi r_{H}^{2} n = \nu \ll 1$ ($n$ is the density of
magnetoexcitons at $T = 0$) (see Ref[\onlinecite{Musin}]):

\begin{eqnarray}
\label{muelhol}
 \mu _{e} = \mu _{h} = E_{0} \nu      .
\end{eqnarray}

and the chemical potential of magnetoexcitons is

\begin{eqnarray}
\label{chem}
\mu _{ex}  =   - T ln(\frac{m_{H}T}{2\pi \hbar ^{2} n_{ex}})
+ 2 T (\frac{\pi \hbar ^{2} n_{ex}}{m_{H}} +
 \pi T^{-2/3}(eD)^{4/3} n_{ex} ln(\frac{2\pi \hbar ^{2} n_{ex}}{m_{H}T}))
 + E_{0} ,
\end{eqnarray}

where $\nu $ is the filling factor and
$E_{0}$ is the magnetoexciton energy; at high
magnetic fields $E_{0} = \sqrt{\frac{\pi }{2}}\frac{e^{2}}{r_{H}}$
at $D \ll a(H,D)$ and
$E_{0} = \frac{e^{2}}{D}$
at $D \gg a(H,D)$. \cite{Ruvinskiy}

We obtain the equation for
the characteristic temperature of ionization $T_{i}(n_{ex},D,H)$
 from   the condition of the ionization equilibrium: \cite{Landaul}

\begin{eqnarray}
\label{ioeq}
0 = \mu _{ex} - \mu _{e} - \mu _{h} =  - T ln(\frac{m_{H}T}{2\pi \hbar ^{2}
n_{ex}}) +  2 T (\frac{\pi \hbar ^{2} n_{ex}}{2m_{H}T} +  \nonumber \\
+  \pi T^{-2/3}(eD)^{4/3} n_{ex} ln(\frac{2\pi \hbar ^{2} n_{ex}}{m_{H}T})
 + E_{0} - E_{0} \nu .
\end{eqnarray}

As $n_{ex} \to 0$ the characteristic temperature of ionization
 $T_{i}(n_{ex},D,H) \to 0$ (and the same for antiexcitons).
 The dependence of maximal $T_{i}(H)$ (corresponding to the maximal
magnetoexciton density)  is defined by the magnetoexciton
binding energy $E_{0}$ (so $T_{i} \sim \sqrt{H}$ at small $D$)
 and rises {\it vs.} magnetic field and decreases {\it vs.} the interlayer
 separation increase.

If we introduce $y = \frac{E_{deg}}{E_{0}}$ ($E_{deg} =
\frac{2\pi \hbar ^{2} n_{ex}}{m_{H}}$ is  the energy of degeneration)
 and $x = \frac{T}{E_{0}}$, this dependence is shown on Fig.3.

{\bf b. Quantum transition to two-layer incompressible liquid state.}

 The rare excitonic  system    is stable
 at $D < D_{cr}(H)$ and $T = 0$ when the magnetoexciton energy $E_{exc}(D,H)$
(calculated in Ref.[\onlinecite{Ruvinskiy}]) is larger than
the sum of activation energies $E_{L} = k\frac{e^{2}}{\epsilon r_{H}}$ for
 incompressible
Laughlin
  liquids of electrons or holes;  $k = 0.06$
for $\nu = \frac{1}{3}$ etc. \cite{Hall} (compare Ref.[\onlinecite{Yoshioka}]
 for  stability of {\it dense} excitonic phase --- see below).
Since $k \ll 1$, the critical value $D_{cr} \gg r_{H}$.
In this case one has $E_{exc} =
\frac{e^{2}}{\epsilon D}(1 - \frac{r_{H}^{2}}{D^{2}})$
for a magnetoexciton with quantum numbers $n = m = 0$
(see Ref.[\onlinecite{Ruvinskiy}]).
As a  result we have from the stability condition (see above)
 $D_{cr} = r_{H} (\frac{1}{2k} - 2k)$.
 For greater $\nu $  it gives an upper bound on $D_{cr}$.
  In the rare system in high magnetic
 fields $\mu \ll E_{ex}$.
The coefficient $k$ in the activation energy $E_{L}$ may be represented
as $k = k_{0}\sqrt{\nu }$. So from the relation between $D_{cr}$ and $r_{H}$
one has: $\nu _{cr} = \frac{1}{k_{0}^{2}} \frac{r_{H}^{2}}{D^{2}}
(1 - \frac{r_{H}^{2}}{8D^{2}})$.
Thus maximal density for stable magnetoexciton phase is
$n_{max} = \nu _{cr}/2\pi r_{H}^{2}$ (see below).
Hence, the {\it maximum} Kosterlitz-Thouless temperature,
at which  superfluidity appeares in the system
is $T_{c}^{max} \sim n^{max}(H,D)/m_{H} \sim
 {\sqrt H}$ at $D \le r_{H}$ or $T_{c}^{max} \sim H^{-1}$  at $D \gg r_{H}$
 in high magnetic fields.  It would be  interesting to check this
 fact in experiments on magnetoexciton systems.
Note that if at a given density of e and h and a given magnetic field $H$
several Landau levels are filled (but high field limit
$r_{H} \ll a_{0}^{*}$ is true) the
superfluid phase can exist for magnetoexcitons
on the highest nonfilled Landau level.
At $\nu = \frac{1}{2}$ the elecron-hole phase  can be unstable  due to
the pairing of electron and hole composite
fermions (which form the Fermi surface of composite
fermions in the mean field approximation \cite{HLR}).

\section{Phase transitions in the "dense" system}

 In the "dense" system (at $\nu \sim 1/2$ contrary to rare system at
$\nu \ll 1$) the ionization of  magnetoexcitons
 can be estimated in the Gor'kov approximation. \cite{Musin}
 It takes place at the temperature
$T_{i}^{dense}$:

\begin{eqnarray}
\label{mus1}
T_{i}^{dense} = \frac{J}{2}\frac{1 - 2\nu }{ln(\nu ^{-1} - 1)}   ,
\end{eqnarray}

where

\begin{eqnarray}
\label{mus2}
J = \int_{}^{} \frac{d^{2}q}{(2\pi )^{2}} V_{12}(q) exp(-q^{2}r_{H}^{2}/2)
\end{eqnarray}

and

\begin{eqnarray}
\label{mus3}
V_{12}(q) =  \frac{2\pi e^{2}}{q} exp(-qD) .
\end{eqnarray}

So the temperature of the ionization crossover
 in the Gor'kov approximation increases with rise of
magnetic field as $\sqrt{H}$ and decreases with the interlayer separation
 (it is consist and with the estimation for the rare system ---
see above).

Let us calculate now  the density of the normal component $n_{n}$.
The contribution of the one-particle overgap excitations
to the density of the normal component is determined by the transitions
between Landau levels. \cite{Lerner} So in high magnetic fields
this contribution is negligible as
 $\frac{e^{2}}{r_{H}\omega _{c}} \ll 1$.  We need also  to take
account of the contribution of collective excitations to the density of
the normal component. In contrast to superconductors, where, as a consequence
of the charge of the Cooper pairs, instead of an acoustic spectrum of
collective oscillations high-frequent plasma mode
arises, in the exciton phase
$e-h$ pairs are neutral and zero-gap  collective excitations
 exist. At low temperatures
the contribution of the elementary excitations in thermodynamic equilibrium
can be described in the approximation of an ideal Bose gas.
So the density of the superfluid component $n_{s}$  can be estimated
by Eq.(\ref{n_s}). Now one need to estimate the sound velocity $c_{s}$
of collective mode
in the system.

It can be shown easily that the
equations of the hydrodynamics  of magnetoexcitons contain
${\bf P'} = \frac{M}{m_{H}} {\bf P}$
 instead  of ${\bf P}$.
  So we have for the sound
velocity:

\begin{eqnarray}
\label{sd}
c_{s}^{2} =  \frac{\partial P'}{\partial \rho }  =
\frac{M}{m_{H}} \frac{\partial P}{\partial \rho }  .
\end{eqnarray}

In result we obtain the expression for
sound velocity with magnetic mass $m_{H}$ instead of $M$
(compare Ref.[\onlinecite{Abrikosov}]):

\begin{eqnarray}
\label{sd1}
c_{s}^{2} =  \frac{n}{m_{H}} \frac{\partial \mu }{\partial n }  .
\end{eqnarray}

Using the relation $\nu = 2\pi r_{H}^{2} n$  we have:

\begin{eqnarray}
\label{sd2}
c_{s} =  \sqrt{\frac{\nu }{m_{H}} \frac{\partial \mu }{\partial \nu }}  .
\end{eqnarray}

The chemical potential of dense system of
electron-hole pairs with spatially separated $e$ and $h$
in high magnetic field is \cite{Musin}:

\begin{eqnarray}
\label{mudense}
\mu = - J + 2\nu (I - J)  .
\end{eqnarray}

Here $I = \int_{}^{} \frac{d^{2}q}{(2\pi )^{2}} V_{11}(q)
exp(-q^{2}r_{H}^{2}/2)$,
$V_{11}(q) = \frac{2\pi e^{2}}{q}$, and $J$ is defined by
Eq.(\ref{mus2}).

From Eqs.(\ref{mudense}) and (\ref{sd2}), one has for the sound
velocity

\begin{eqnarray}
\label{sd3}
c_{s} =  \sqrt{\frac{2\nu (I - J)}{m_{H}}}  .
\end{eqnarray}

The temperature of the Kosterlitz-Thouless  transition to the superfluid
state can be calculated by using Eq.(\ref{tct}) (see Sec.V), where
 $T_{c}^{0}$ is the auxiliary quantity equal to the temperature  of
vanishing of  superfluid density in the mean field approximation
 $n_{s}(T_{c}^{0}) = 0$:

\begin{eqnarray}
\label{tct00}
T_c^0 = \left( \frac{\nu c_s^4 m_H}{3 r_{H}^{2} \zeta (3)} \right)^{1/3} =
\left( \frac{32 r_{H}^{4} (I - J)^{2} }{3 \zeta (3) m_{H}} \right)^{1/3}
\frac{\nu }{2 r_{H}^{2}} .
\end{eqnarray}

In high magnetic fields for dense system
the Kosterlitz-Thouless temperature is obtained
by substituting of $T_c^0$ from Eq.(\ref{tct00}) to Eq.(\ref{tct}).
 Kosterlitz-Thouless temperature decreases {\it vs.}  magnetic field
analogously to the rare system  (see above).

Then we have:

\begin{eqnarray}
\label{diff}
I - J = e^{2} \int_{0}^{\infty} (1 - e^{-qD})
 e^{\frac{-q^{2}r_{H}^{2}}{2}} dq.
\end{eqnarray}

Let us consider for estimate
small interlayer distances  $D \ll 1/q_{0}$, where $q_{0}$
is the characteristic wave vector $q_{0} \sim 1/r_{0}$,
where $r_{0} = 1/\sqrt{\pi n_{ex}}$  is the mean distance between particles.
Maximal density of particles at $\nu = \frac{1}{2}$ is $n_{ex} =
\frac{1}{4\pi r_{H}^{2}}$
(at $\nu > 1/2$ we deal with antiexcitons - see below).
If $D \lesssim r_{H}$, using Eq.(\ref{diff}), we have:

\begin{eqnarray}
\label{diffe}
I - J = e^{2} D \int_{0}^{\infty} q e^{\frac{-q^{2}r_{H}^{2}}{2}} dq =
\frac{e^{2} D}{r_{H}^{2}}.
\end{eqnarray}

Then we have for the temperature of the phase transition to the superfluid
state in the mean field approximation at small $D$:

\begin{eqnarray}
\label{tct00sm}
T_c^0 = \left( \frac{2 \pi n_{ex} c_s^4 m_H}{3 \zeta (3)} \right)^{1/3} =
\left( \frac{32 e^{4} D^{2}}{3 \zeta (3) m_{H}} \right)^{1/3}
\pi n_{ex}  .
\end{eqnarray}

As we mentioned above in high magnetic fields
the symmetry $\nu \to 1 - \nu $,  $e \longleftrightarrow h$ takes place
at the Landau level and unoccupied states on
Landau levels for spatially separated electrons and holes can bind
to form "antiexcitons" and superfluidity of "antiexcitons"  can  take place.
 The Kosterlitz-Thouless temperature  for
superfluidity of
 antiexcitons as function of
 $H$, $D$  is symmetrical to that for excitons. Hence,
 in expressions for temperatures of phase transitions in Eqs.(\ref{tct00}),
and (\ref{tct00sm}) we can
use $\nu (1 - \nu )$  instead of $\nu$.

 We can see
that the temperature of the Kosterlitz-Thouless transition to the superfluid
state at fixed density
decrease with rise of magnetic field as $H^{-1/6}$ (for the rare system
as $H^{-1/2}$ --- see above). $T_{c}$ decreases also {\it vs.}
interlayer separation.
At values of $D$  greater than some critical one $D_{cr}$ the
 dense superfluid
system of magnetoexitons must transform into the system of two
incompressible liquids. \cite{Yoshioka}
 The dense excitonic  system    is stable
 at $D < D_{cr}(H)$ and $T = 0$ when the Hartree-Fock energy
  is larger than
the sum of activation energies $E_{L} = k\frac{e^{2}}{\epsilon r_{H}}$ for
 incompressible  Laughlin
  liquids of electrons or holes (it is consistent with the
results for rare system ---
see above).

\section{Magnetoexcitons in unbalanced two-layer electron system}

The system of indirect magnetoexcitons  can appear also
in  unbalanced two-layer {\it electron} system  in  CQW in strong
magnetic fields near the filling factor $\nu = 1$. An external electric
voltage between  quantum wells change the filling, so say in the first
 quantum well the filling factor will be
$\nu _{1} = \Delta \nu \ll \frac{1}{2}$ and
in another one it will be $\nu _{2} = 1 - \Delta \nu $.
Unbalanced
filling factors
 $\nu _{1} = 1 + \Delta \nu $,
 $\nu _{2} = 1 - \Delta \nu $ is also possible.
  Thus in the first
quantum well (QW) there are rare electrons on the second Landau level,
and in the second QW there are rare empty places
("holes") on the first Landau level.
  "Excess" electrons in the first QW and "holes" in the second
QW can bound to indirect magnetoexcitons
with the density  $n_{ex} =
 eH\Delta \nu /2\pi $.   Superfluidity in two-layer  $e-e$ system
in high magnetic fields in the cases mentioned is analogous to the
superfluidity of two-layer $e-h$ system.

 The expressions for critical values
$\nu _{cr}$, $D_{cr}$  for Kosterlitz-Thouless temperature $T_{c}$
calculated above are
applicable also for two-layer electron system under consideration.
So, in this approximation
 the phase diagram for unbalanced
two-layer electron system  is analogous to the
two-layer electron-hole system.
In condition for the stability of superfluid magnetoexciton liquid
relating to quantum
 transition to two incompressible  Laughlin liquids (see Sec. VII)
one must use $\Delta \nu $  instead of $\nu $.
Due to Jain's mapping of fractional Landau level fillings to integer Landau
level fillings \cite{Jain}
analogous results take place also for slightly unbalanced
fractional fillings.

\section {Possible experimental manifestations of magnetoexciton
superfluidity}

The appearance of local superfluid density above $T_{c}$
  can be manifested, for
example, in observations of temperature dependence of the
 exciton diffusion on intermediate distances
(with the help of local measurements of exciton photoluminescence at
two points using optical fibers or pinholes in experiments like those
in Ref.[\onlinecite{Zrenner}]).

 The superfluid
state at $T < T_{c}$ can manifested itself in the existence of persistent
("superconducting") oppositely directed electric currents in each layer.
The interlayer tunneling in
 an {\it equilibrium} spatially separated electron-hole system
 leads to interesting
Josephson phenomena in the system: to a transverse Josephson current,
inhomogeneous (many sin-Gordon soliton) longitudinal
  currents, \cite{Klyuch}  diamagnetism
in a magnetic field $H$ parallel to the junction (when $H$
 is less than some critical value $H_{c1}$, depending on the
tunneling coefficient), and a mixed state with Josephson vortices for
$H > H_{C1}$ \cite{yel} (in addition, taking tunneling into account
leads to a loss of symmetry of the order parameter and to a change
in the character of the phase transition).

\section{ Conclusions }

 We have shown that
at fixed exciton density $n_{ex}$ the  Kosterlitz-Thouless
temperature $T_{c}$ for the onset of superfluidity of magnetoexcitons
 decreases as a function of magnetic field as $H^{-\frac{1}{2}}$
(at $D \lesssim r_{H}$). But
{\it the maximal} $T_{c}$
(corresponding to the maximal magnetoexciton densities)
increases with $H$ in high magnetic fields
 as $T_{c}^{max}(H,D) \sim {\sqrt H}$ (at $D \lesssim r_{H}$).
 This fact needs to be compared in detail with the results of experimental
studies of the collective properties of magnetoexcitons.
The excitonic
 phase is more stable than the Laughlin states of electrons and holes
 at a given  Landau filling
$\nu $ if $D < D_{cr} = r_{H}
(\frac{1}{2k} - 2k)$, where $k$ is the coefficient in the Laughlin activation
energy. Below the Kosterlitz-Thouless temperature
one can observe the appearance of persistent currents in separate
quantum wells.
We have shown, that in extremely high magnetic fields the
 system of indirect magnetoexcitons at fixed $T$
has the statistical
properties of almost ideal gas.
At small interlayer distances $D \lesssim r_{H}$
the temperature of the Kosterlitz-Thouless transition to the superfluid
state in the dense system decreases  with rise of magnetic field as
$H^{-1/6}$
due to rise of the magnetic mass of indirect
 magnetoexciton. \cite{Ruvinskiy}
We discuss also the quantum transition to the incompressible liquid state.
We calculated the characteristic temperature of insulator-metal transition
 and established that it rises with magnetic field and
 decreases with interlayer  separation.
Note that in some region of Landau filling inside ($0,\nu _{cr}$)
(see above) crystal phase
of indirect magnetoexcitons must exist (see Ref.[\onlinecite{Quinn}]).
Note that its melting curve is analogous to one for electron crystal
near metal gate because due to image forces effective interaction in
 the system
is also dipole-dipole interaction \cite{Ab}$^,$\cite{cr}
(the difference in results for these
two systems is due to their different statistics).

Yu.E.L. is grateful to J.K.Jain, A.MacDonald and G.Vignale for interesting
discussions.
The work was supported by Russian Foundation of Basic Research, INTAS
and Program "Physics of Solid Nanostructures".
O.L.B. was supported by the Program "Soros PhD students" and
ICFPM (International Center for Fundamental Physics in Moscow)
Fellowship Program 1998.

\newpage

%\begin{large}

\begin{center}
{\bf Yu.E.Lozovik, O.L.Berman, V.G.Tsvetus}
\end{center}

\begin{center}
{\bf Captures to Figures}
\end{center}

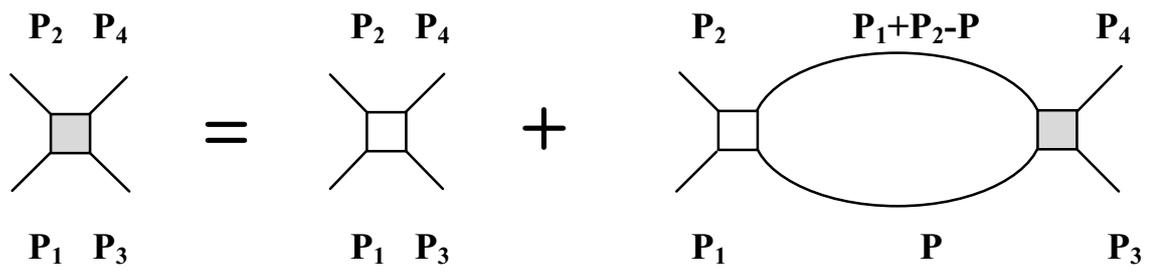
Fig.1. The equation for the vertex $\Gamma $ in
the representation of magnetic momenta
${\bf P}$  and quantum numbers $m$ and $n$

Fig.2.  Dependence of temperature of Kosterlitz-Thouless transition
$T_{c}$
on magnetic field
$H$ at  different  of  interwell separations $D$

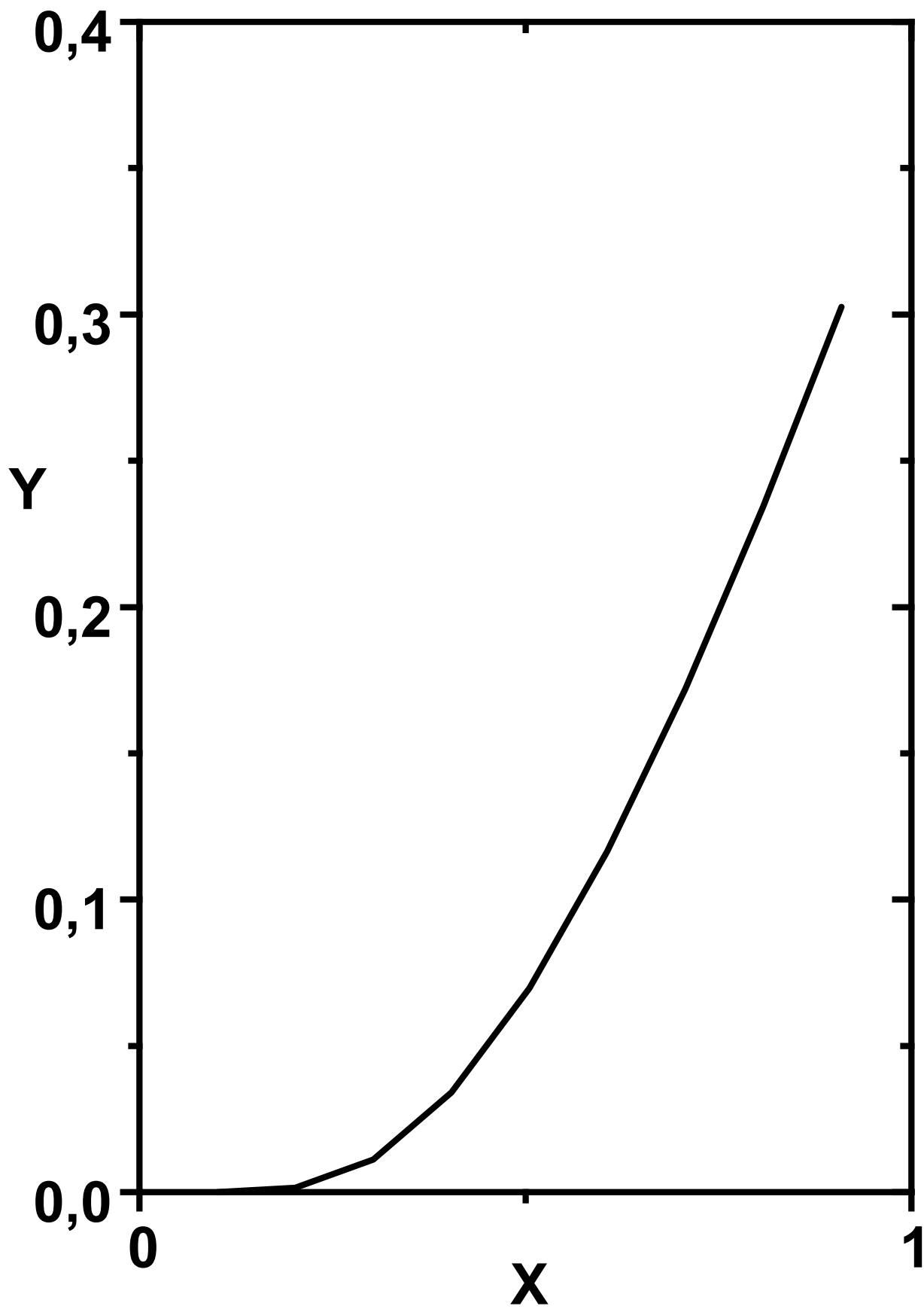
Fig.3. Magnetoexcitons existence line $T_{i}(H,n)$;
$y = \frac{E_{deg}}{E_{0}}$, $E_{deg} =
\frac{2\pi \hbar ^{2} n_{ex}}{m_{H}}$ is the energy of degeneration
 and $x = \frac{T}{E_{0}}$.

%\end{large}

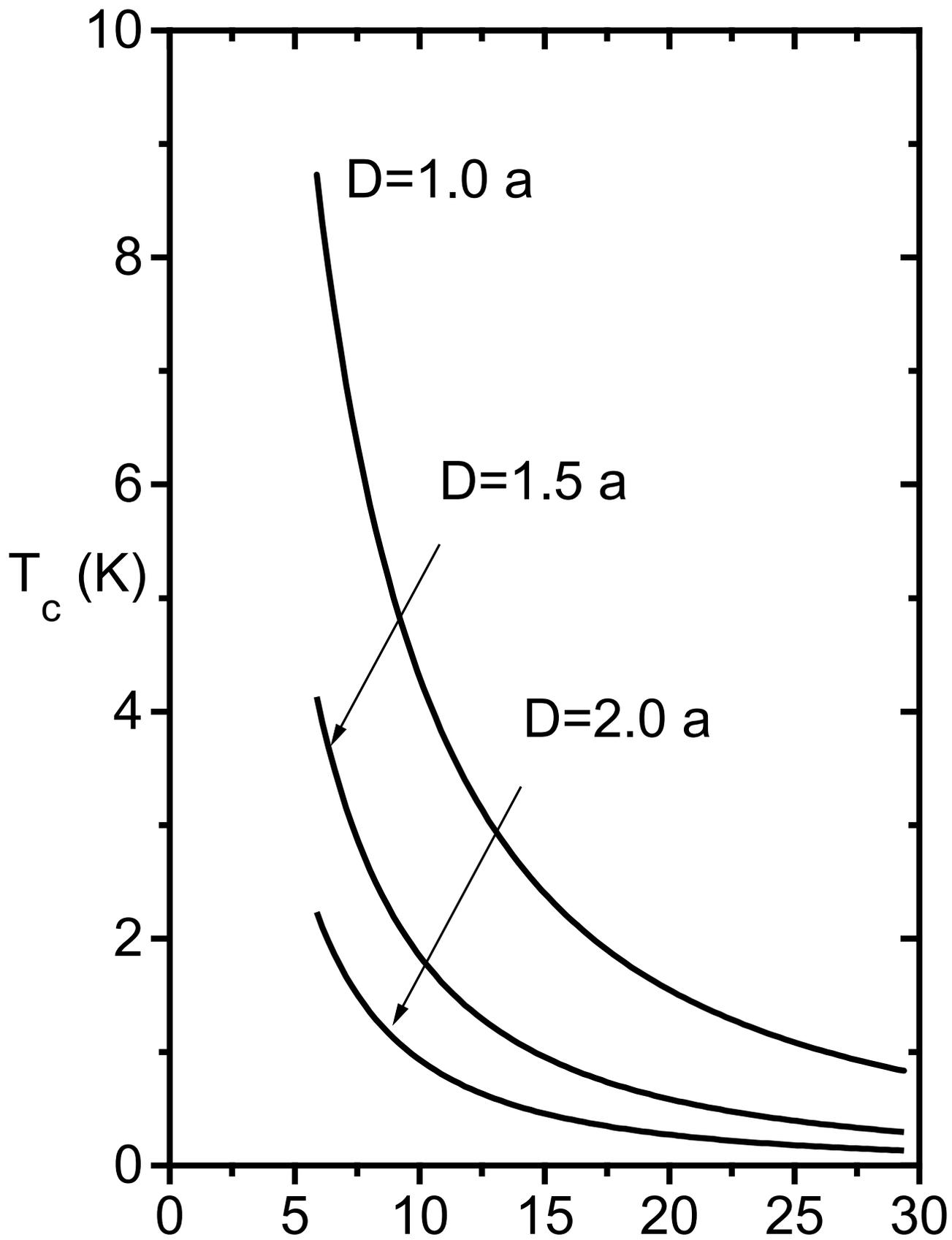
\end{document}